\documentclass[12pt,preprint]{emulateapj}
\usepackage{natbib}
\usepackage{aas_macros}
\DeclareMathAlphabet{\mathsc}{OT1}{cmr}{m}{sc}

\newcommand{\OII}{[O {\sc ii}]}

\begin{document}

\title{
Environmental Effects on the Star Formation Activity at $z\sim 0.9$ in the COSMOS Field\altaffilmark{*}}

\author{ 
M. Kajisawa\altaffilmark{1,2},
Y. Shioya\altaffilmark{1},
Y. Aida\altaffilmark{2}, 
Y. Ideue\altaffilmark{2},
Y. Taniguchi\altaffilmark{1},
T. Nagao\altaffilmark{1,3,4},
T. Murayama        \altaffilmark{5},
K. Matsubayashi\altaffilmark{1}, 
L. Riguccini\altaffilmark{1,6}
}

\altaffiltext{*}{Based on observations with the NASA/ESA 
        {\it Hubble Space Telescope}, obtained at the Space Telescope Science 
	Institute, which is operated by AURA Inc, under NASA contract NAS 
	5-26555. Also based on observations made with the Spitzer Space 
	Telescope, which is operated by the 
	Jet Propulsion Laboratory, California Institute of Technology, 
	under NASA contract 1407. Also based on data collected at;  
	the Subaru Telescope, which is operated by the National Astronomical 
	Observatory of Japan; the XMM-Newton, an ESA science mission with 
	instruments and contributions directly funded by ESA Member States and
	NASA; the European Southern Observatory under Large 
	Program 175.A-0839, Chile; Kitt Peak National Observatory, Cerro 
	Tololo Inter-American Observatory and the National Optical Astronomy 
	Observatory, which are operated by the Association of Universities for 
	Research in Astronomy, Inc. (AURA) under cooperative agreement with 
	the National Science Foundation; and the Canada-France-Hawaii 
	Telescope with MegaPrime/MegaCam operated as a joint project by the 
	CFHT Corporation, CEA/DAPNIA, the NRC and CADC of Canada, the CNRS 
	of France, TERAPIX and the Univ. of Hawaii.}

\altaffiltext{1}{Research Center for Space and Cosmic Evolution, 
        Ehime University, Bunkyo-cho, Matsuyama 790-8577, Japan 
       {\it e-mail kajisawa@cosmos.phys.sci.ehime-u.ac.jp}}
\altaffiltext{2}{Graduate School of Science and Engineering, Ehime University, 
        Bunkyo-cho, Matsuyama 790-8577, Japan}
\altaffiltext{3}{The Hakubi Project, Kyoto University, Yoshida-Ushinomiya-cho,
Sakyo-ku, Kyoto 606-8302, Japan}
\altaffiltext{4}{Department of Astronomy, Kyoto University, Kitashirakawa-Oiwake-cho,
Sakyo-ku, Kyoto 606-8502, Japan}
\altaffiltext{5}{Astronomical Institute, Graduate School of Science,
        Tohoku University, Aramaki, Aoba, Sendai 980-8578, Japan}
\altaffiltext{6}{NASA Ames Research Center, Moffett Field, CA 94035}

\shortauthors{M. Kajisawa et al.}
\shorttitle{\OII\ Emitters at $z\simeq 0.9$}

\begin{abstract}
We investigated the fraction of \OII\ emitters in galaxies 
at $z\sim0.9$ as a function of the local galaxy density
in the Hubble Space Telescope (HST) COSMOS 2 square degree field. 
\OII\ emitters are selected by the narrow-band excess technique with 
the NB711-band imaging data taken with Suprime-Cam on the Subaru telescope. 
We carefully selected 614 photo-z selected galaxies with 
$M_{U3500} < -19.31$ at $z=0.901$ -- 0.920, 
which includes 195 \OII\ emitters, to directly compare results with our previous 
study at $z\sim 1.2$. 
We found that the fraction is almost constant 
at $0.3 \; {\rm Mpc^{-2}} < \Sigma_{\rm 10th} < 10 \; {\rm Mpc^{-2}}$. 
We also checked the fraction of galaxies with blue rest-frame colors 
of $NUV-R<2$ in our photo-z selected sample, and found that 
the fraction of blue galaxies does not significantly depend on the local density.
On the other hand, 
the semi-analytic model of galaxy formation predicted that the fraction of 
star-forming galaxies at $z\sim0.9$ decreases with increasing the projected galaxy density 
even if the effects of the projection and the photo-z error in our analysis were taken 
into account. 
The fraction of \OII\ emitters decreases from $\sim 60\%$ at $z\sim1.2$ to 
$\sim 30\%$ at $z\sim 0.9$ independent of the galaxy environment. 
The decrease of the \OII\ emitter fraction could be explained mainly by the rapid decrease 
of the star formation activity in the universe from $z \sim 1.2$ to $z \sim 0.9$. 
\end{abstract}

\keywords{galaxies: environment --- 
          galaxies: evolution --- 
          galaxies: star formation}


\section{INTRODUCTION}

It is known that star formation activity in galaxies strongly depends on environment in 
the present universe.
The high-density regions such as clusters of galaxies are dominated by passively evolving 
early-type galaxies, while there are many star-forming late-type galaxies in field 
(low-density) environments (e.g., \citealp{dre80}; \citealp{got03}; \citealp{bam09}).
The fraction of star-forming galaxies systematically decreases with increasing local 
galaxy density (e.g., \citealp{gom03}; \citealp{bal04}; \citealp{kau04}; \citealp{tan04}).
From these findings, it is considered that the star formation history of
galaxies in general depends on environment.

In order to understand how the star formation history of galaxies depends on environment, 
it is important to investigate the star formation activity  of galaxies as a function 
of environment in the early universe.
Several such environmental studies have been carried out at $z>1$, when the cosmic star 
formation rate density reached its peak.
\cite{hay10} studied the fraction of narrow-band selected \OII\ emitters as a 
function of the local galaxy density around a cluster at $z\sim1.5$, and found that the 
fraction of such star-forming galaxies is nearly independent of the local density, and 
does not decrease even in the core of the cluster.
\cite{tra10} also found that the fraction of actively star-forming galaxies with 
 bright IR luminosity slightly increases with the local galaxy density in a cluster 
at $z=1.62$.
\cite{ide09} similarly investigated the fraction of \OII\ emitters in more general 
environments at $z\sim1.2$, and found that the fraction is almost constant from low-density 
to medium-density environments.
These results suggest that the relation between the star formation activity and the galaxy 
environment changed between $z>1$ and $z\sim0$.
Since the cosmic star formation rate (SFR) density decreases from $z\sim1$ to the present by 
about an order of magnitude (e.g., \citealp{hop06}; \citealp{shi08}; \citealp{wes10}), 
the change 
of the environmental dependence of the star formation activity might be directly related 
with the decrease of the global SFR density in the universe.
Therefore, it is interesting to investigate the star formation activity as a function 
of environment in detail at $z\sim1$, when the cosmic SFR density started 
to decrease.

At $z\sim1$, several studies in general environments claimed that the average star 
formation rate of galaxies increases with the local galaxy density (e.g., \citealp{elb07}; \citealp{coo08}), while some studies around clusters of galaxies reported that 
the fraction of star-forming galaxies decreases with the local density (e.g., \citealp{pog08}; \citealp{pat09}; \citealp{koy10}; \citealp{pat11}).
Recently, \cite{sob11} carried out a wide-field near-infrared narrow-band survey 
in the COSMOS and UKIDSS UDS fields.
They found that the fraction of narrow-band selected H$\alpha$ emitters is nearly constant 
or slightly increases with the local galaxy density from low-density to medium-density 
environments, and then the fraction decreases towards the highest-density regions such 
as rich clusters, which is consistent with previous studies both in general fields and 
clusters regions.
Such somewhat complicated environmental dependence of the star formation activity might 
be considered to be intermediate between those at $z>1$ and at the present.

The next important step is to investigate the evolution of the star formation activity 
as a function of environment by directly comparing those in different epochs.
For example, \cite{elb07} and \cite{coo08} made comparisons between $z\sim1$ and $z\sim0$ 
in general fields, while \cite{pog08} compared the results in groups and 
clusters at $z=$ 0.4--0.8 with those at $z\sim0$. 
In this paper, we focus on the evolution between $z\sim1.2$ and $z\sim0.9$, when 
the cosmic SFR density began to decrease. 
Using the optical narrow-band imaging 
data in the COSMOS survey \citep{sco07} obtained with Suprime-Cam on the Subaru 
Telescope, 
we investigated the fraction of narrow-band selected \OII\ emitters in galaxies at 
$z\sim0.9$ as a function of the local galaxy density. 
The combination of the wide area of the COSMOS survey and the \OII\ emitter selection 
allows us to construct a large sample of star-forming galaxies with a secure redshift 
identification. 
By carefully choosing the selection criteria for our sample and adopting the same method 
for the estimate of the local galaxy density as our previous study at $z\sim1.2$ in the 
COSMOS field \citep{ide09}, we directly compare the star formation activities 
as a function of environment between $z\sim 0.9$ and  $z\sim1.2$.
Section 2 describes the data and the methods for the sample selection and the environment 
estimate.
In Section 3, we show the fraction of \OII\ emitters in galaxies 
 at $z\sim0.9$ as a function of the environment and compare it with that at $z\sim1.2$, 
and then check the robustness of the results. 
In Section 4, we compare our results with previous studies, and then 
discuss the evolution of the fraction in the redshift interval and 
its relation to the decrease of the star formation activity in the universe.

Throughout this paper, magnitudes are given in the AB system. We adopt a flat universe 
with $\Omega_{\rm matter}=0.3$, $\Omega_{\Lambda}=0.7$, and $H_{0}=70$ km 
s$^{-1}$ Mpc$^{-1}$.

\section{SAMPLE AND ANALYSIS}

\subsection{Samples}
In this study, we use a sample of galaxies with photometric redshifts of 
$z=0.901$ -- 0.920 from the COSMOS photometric redshift catalog \citep{ilb09}.
We can select \OII\ emitters using the NB711 narrow-band data for the redshift interval.  
In order to directly compare with our previous results at $z\sim1.2$ (\citealp{ide09},
hereafter I09), we need to match the magnitude limit of the sample to that at $z\sim1.2$. 
In I09, we used the sample of galaxies with $i^\prime < 24$ at $z\sim1.2$, 
which corresponds to the rest-frame 3500 \AA~ absolute magnitude of 
$M_{U3500}<-19.71$. Therefore we constructed a sample of galaxies 
with $M_{U3500}<-19.71$ at $z=0.901$ -- 0.920 (hereafter Sample A). 
The rest-frame 3500 \AA~ absolute magnitude was calculated from 
the best-fit SED template derived in the photo-$z$ calculation \citep{ilb09} 
for each galaxy. 

In addition to the Sample A, we also construct another sample of galaxies 
at $z=0.901$ -- 0.920 using the different magnitude limit, for which the luminosity 
evolution at the rest-frame 3500 \AA~ between $z\sim1.2$ and $z\sim0.9$ 
is taken into account. 
In order to measure the strength of the luminosity evolution, we derived 
the rest-frame 3500 \AA~ luminosity functions (LFs) for the $z\sim0.9$ and $z\sim1.2$ 
samples, and fitted these LFs with the Schechter function.    
The estimated Schechter parameters are 
$M_\star = -19.55$ and $\log \phi_\star = -2.01$ for the $z\sim0.9$ sample 
and $M_\star = -19.95$ and $\log \phi_\star = -1.94$ for the $z\sim1.2$ sample. 
The faint-end slope was fixed to $\alpha = -1.0$ in the fitting for the both redshifts.
Therefore we assumed that galaxies become fainter by 0.4 mag at the rest-frame 3500 \AA~ 
from $z\sim1.2$ to $z\sim0.9$, and used $-19.71 + 0.4 = -19.31$ as another magnitude limit.
The sample of galaxies with $M_{U3500}< -19.31$  at $z=0.901$ -- 0.920
 is referred to as Sample B.

\begin{figure}[t]
\begin{center}
\includegraphics[width=80mm]{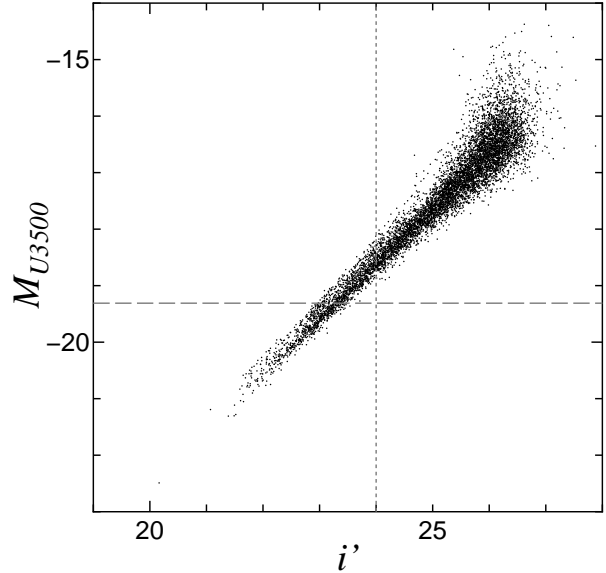}
\caption{$M_{U3500}$ vs. $i^\prime$ for galaxies 
at $0.901 \le z_{\rm phot} \le 0.920$ from the COSMOS photometric redshift 
catalog. 
Horizontal dashed line and vertical dotted line show 
$M_{U3500}=-19.31$ and $i^\prime=24.0$, respectively.
All galaxies with $M_{U3500}<-19.31$ satisfy $i^\prime<24.0$. 
}
\label{fig:m3500i}
\end{center}
\end{figure}
For the redshift range, we can almost completely sample galaxies even with 
 $M_{U3500}< -19.31$ (Figure 1). All galaxies in the samples have $i^\prime < 24$, 
and the photometric redshift accuracy is the same as in our previous 
study at $z\sim1.2$. The numbers of galaxies in the samples are 373 for the Sample A and 
733 for the Sample B. The effective survey area is 5540 arcmin$^2$ and 
the redshift interval of  $z=0.901$ -- 0.920 corresponds to the co-moving depth of 
50 Mpc. Then our effective survey volume is $2.28\times 10^{5}$ Mpc$^3$ 
given the assumed cosmology.

\subsection{[O {\sc ii}] Emitter Selection}

We select \OII\ emitters from the samples mentioned above as star-forming galaxies 
using the narrow-band excess technique. 
In order to select the \OII\ emitter, 
we use the $r^{\prime}$, $i^{\prime}$, and NB711 bands photometry from 
the COSMOS photometric catalog \citep{cap07}. NB711 is a narrow band with 
a central wavelength of 7119.6 \AA~ and a full width at half maximum of 72.5 \AA , 
which covers the redshifted \OII $\lambda\lambda$3727 emission lines for galaxies at 
  $z=0.901$ -- 0.920. We used the  $3^{\prime \prime}$ diameter aperture magnitudes in 
three bands. 
We adopted the correction for the photometric zero point presented by 
\cite{ilb09}, which is calculated by comparing the observed multi broad-band 
photometry for galaxies with spectroscopic identification with the best-fit model templates.
The zero-point corrections are 0.003, 0.019, and 0.014 mag for $r^{\prime }$, $i^{\prime }$, 
and NB711 bands, respectively.
The limiting magnitudes are $r^{\prime}_{\rm lim} = 26.6$,
$i^{\prime}_{\rm lim} = 26.1$, and ${\rm NB711}_{\rm lim} = 24.6$, for a $3 \sigma$ detection 
on a $3^{\prime \prime}\phi$ diameter aperture. 
It is noted that we use the CFHT $i^{*}$-band magnitude instead of 
the Subaru/Suprime-Cam $i^{\prime}$-band one for galaxies 
brighter than $i^{\prime}$ = 21 because such bright galaxies
appear to be affected by the saturation effect in the Suprime-Cam data. 
Details of the imaging data and the photometry are given 
in \cite{tan07} and \cite{cap07}.

In order to select NB711-band excess objects, we calculated a 
  continuum magnitude at the wavelength of NB711 from $r$ and $i$-bands magnitudes as 
$f_{ri} = 0.32f_{r^\prime}+0.68f_{i^\prime}$, where $f_{r^{\prime }}$
and $f_{i^{\prime }}$ are the $r^{\prime }$ and $i^{\prime }$ flux densities,
respectively. Its 3$\sigma$ limiting magnitude is $ri\simeq 26.5$ in a 
$3^{\prime\prime} \phi$ aperture. 
For the bright galaxies with $i^{\prime }< 21$, the
$ri$ continuum is calculated as $f_{ri} = 0.32f_{r^\prime}+0.68f_{i^*}$,
where $f_{i^{*}}$ is the CFHT $i^{*}$ flux density.
\begin{figure}[t]
\begin{center}
\includegraphics[width=80mm]{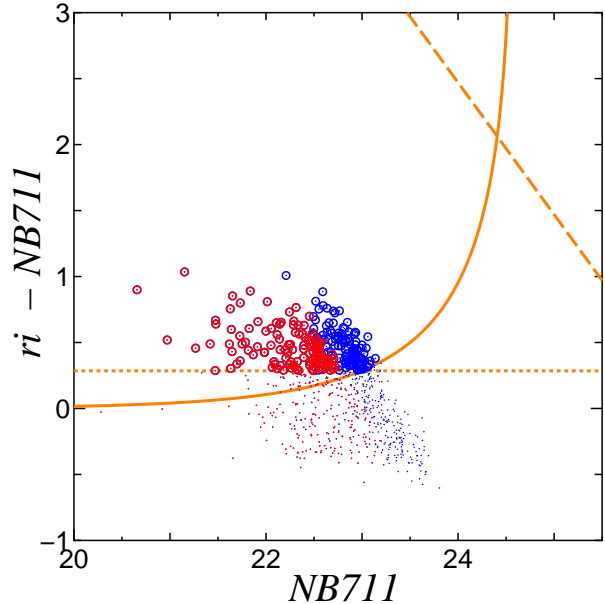}
\caption{$ri-\mathrm{NB}\,711$ vs. $\mathrm{NB}\,711$ for galaxies with 
$M_{U3500} < -19.31$ at $0.901 \le z_{\rm phot} \le 0.920$. 
Large open circles show objects classified as \OII\ emitters. 
Red symbols show galaxies with $M_{U3500} \le -19.71$ (the Sample A), 
while blue symbols represent those with $-19.71 < M_{U3500} < -19.31$ 
(i.e., red $+$ blue $=$ the Sample B).  
The dotted line corresponds to the minimum excess of $ri-\rm NB711=0.285$, and  
the solid curve shows the $3\sigma$ error of $ri-NB711$ color. 
The dashed line represents the $3\sigma$ sensitivity limit for $ri$ magnitude.}
\label{fig:color}
\end{center}
\end{figure}

Then, we select NB711-band excess objects using the following criteria: 
\begin{equation}
ri - NB711 \geq 0.285
\label{eqn:emitter1}
\end{equation}
and
\begin{equation}
ri - NB711 > 3\sigma_{ri - NB711},
\label{eqn:emitter2}
\end{equation}
where
\begin{equation}
3\sigma_{ri - NB711}=-2.5\log \left(1-\sqrt{(f_{3\sigma_{NB711}})^2+
(f_{3\sigma_{ri}})^2}/f_{NB711}\right).
\label{eqn:emitter3}
\end{equation}
The former criterion corresponds to the rest-frame equivalent width 
$EW_0$(\OII) $\geq 12$ \AA, which is the same as that in our previous study 
at $z\sim1.2$ (I09). Figure \ref{fig:color} shows the color-magnitude distribution 
for galaxies at $z=0.901$ -- 0.920. Red symbols represent galaxies with 
$M_{U3500} < -19.71$ (Sample A), and blue ones show galaxies with 
$-19.71 < M_{U3500} < -19.31$. The narrow and broad-bands data 
are deep enough to almost completely sample galaxies with $EW_0$(\OII) $\geq 12$ \AA. 
Here, we exclude X-ray sources as active galactic nuclei (AGNs) 
based on the X-ray information 
given in the COSMOS photo-$z$ catalog \citep{ilb09}.
Finally, we select 118 \OII\ emitters out of 373 galaxies in the Sample A and 
233 \OII\ emitters out of 733 galaxies in the Sample B.

\begin{table*}[t]
\begin{center}
\caption{Summary of the samples}
\label{tb:sample}

\begin{tabular}{ccccc}
\hline\hline
& photometric & limiting & total number & \# of \OII\ emitters  \\
& redshift & magnitude &($\Sigma_{\rm 10th}$ available) &($\Sigma_{\rm 10th}$ available)\\ 
\hline
Sample A & 0.901--0.920 & $M_{U3500} < -19.71$ & 373 (291) & 118 (95) \\ 
Sample B & 0.901--0.920 & $M_{U3500} < -19.31$ & 733 (614) & 233 (195) \\
\hline
\end{tabular}
\end{center}
\end{table*}
We also examine how many AGNs are included 
in our sample using Spitzer IRAC mid-infrared colors. 
\cite{lac04} and \cite{ste05} pointed out that AGNs can be distinguished 
from star-forming galaxies using Spitzer IRAC colors, e.g., $[3.6]-[4.5]$. 
While the ultraviolet to mid-infrared ($\lambda < 5 \; {\rm \mu m}$) continuum 
of star-forming galaxies is the composite stellar continuum that peaks 
at $\sim 1.6 \; {\rm \mu m}$, an AGN continuum is well fit by a power law. 
The infrared colors of AGNs tend to be systematically redder than 
star-forming galaxies. 
As in I09, we select objects with a mid-infrared color of $[3.6] - [4.5] > 0$ as AGN. 
We apply this criterion for the 207 [O {\sc ii}] emitters detected 
in both 3.6 and 4.5 $\rm  \mu m$. 
We find that 2 out of the 207 \OII\ emitters 
in our sample satisfy this criterion;
the fraction of the possible AGNs is 2/207 = 1.0 \% at most. 
Therefore, we consider that the AGN contamination does not affect 
our discussion below.

\subsection{Local Surface Density}
As in I09, we use the 10th nearest neighbor method to estimate 
  the local surface density of galaxies as a measurement of the galaxy environment. 
The projected surface density is calculated as 
\begin{equation}
\Sigma_{\rm 10th}=\frac{11}{\pi r^2} \; ,  
\end{equation}
where $r$ is the distance to 10th nearest neighbor. 
We calculate this distance for galaxies within the redshift $z\pm\sigma_z$  
taking account of the error of the photometric redshift ($\sigma_z = 0.023$ at $z\sim0.9$). 
The redshift slice of $z\pm\sigma_z$ corresponds to the co-moving depth 
of 118 Mpc, while that in our previous study at $z\sim1.2$ is 114 Mpc (I09). 
Therefore, the co-moving depth for the estimate of the projected surface density 
in this study is similar with that in I09.   
We consider that such a small difference in the depth of the redshift slice  
does not affect our results shown in the next section. 
We discard the galaxies near the edge of our field of view, i.e., 
whose $r$ is larger than the distance to the edge of the field. 
This procedure decreases the numbers of galaxies in our samples to 
291 (95 \OII\ emitters) for the Sample A and 614 (195 \OII\ emitters) 
for the Sample B. 
We summarize our samples in Table \ref{tb:sample}.

\section{RESULTS}
\subsection{Fraction of \OII\ emitters at $z\sim0.9$ as a function of local density} 

\begin{figure}[t]
\begin{center}
\includegraphics[width=80mm]{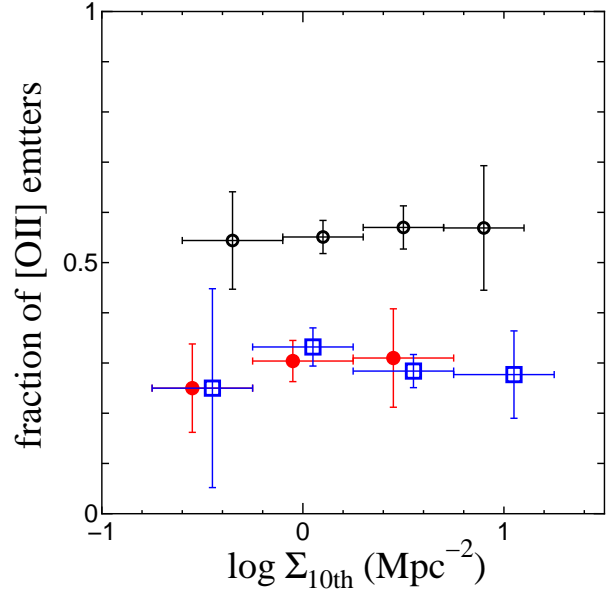} 
\caption{
Fraction of \OII\ emitters as a function of the galaxy local density. 
Filled circles show the result for the Sample A 
and open squares show that for the Sample B. 
Small open circles represent the result at $z \sim 1.2$ 
updated from \cite{ide09} with the newest version of 
the COSMOS photometric redshift catalog.
}
\label{fig:fraction}
\end{center}
\end{figure}

Figure \ref{fig:fraction} shows the fraction of \OII\ emitters in galaxies 
at $z\sim0.9$ as a function of the local galaxy density. 
The results for the Samples A and B are shown as 
solid circles and open squares, respectively. 
For comparison, we also show the results for galaxies at $z\sim1.2$ from I09. 

In Figure \ref{fig:fraction}, 
we cannot find a significant environmental dependence of the fraction of 
 \OII\ emitters in the both Samples A and B. 
The fraction of \OII\ emitters is nearly constant ($\sim 0.3$) for 
between $\Sigma_{\rm 10th} \sim 0.3 \; {\rm Mpc^{-2}}$ and $\sim 10 \; {\rm Mpc^{-2}}$. 
Even if we take the effect of luminosity 
evolution of galaxies into account, the flat distribution holds.
The fraction of \OII\ emitters at $z\sim1.2$ also shows no significant 
environmental dependence, but the fraction is $\sim 0.6$ over a wide range of 
the local density, which is significantly higher than that at $z\sim 0.9$. 
The fraction of \OII\ emitters decreases from $\sim 0.6$ to $\sim 0.3$ between 
$z\sim1.2$ and $z\sim0.9$ in all environments we investigated. 

We check the effects of the projection and the photometric 
redshift error on these results, in the following sections.

\subsection{Fraction of galaxies with blue $NUV - R$ color}  

In the analysis of the previous section, we mainly used the sample of 
galaxies with  $0.901 \le z_{\rm phot} \le 0.920$ (i.e., $\Delta z 
= 0.019$) to select \OII\ emitters as the star-forming population, while 
the local galaxy density are calculated with galaxies within a redshift slice of 
$z\pm \sigma_z$ (i.e., $\Delta z = 0.046$) taking account of the photo-z error.
So the density measurement includes galaxies outside the main sample. 
Furthermore, as we show in Section \ref{sec:specz}, 
\OII\ emitters selected by the NB711-band excess tend to have higher photo-z accuracy
than the other galaxies without the narrow-band excess.
\begin{figure}[t]
\begin{center}
\includegraphics[width=80mm]{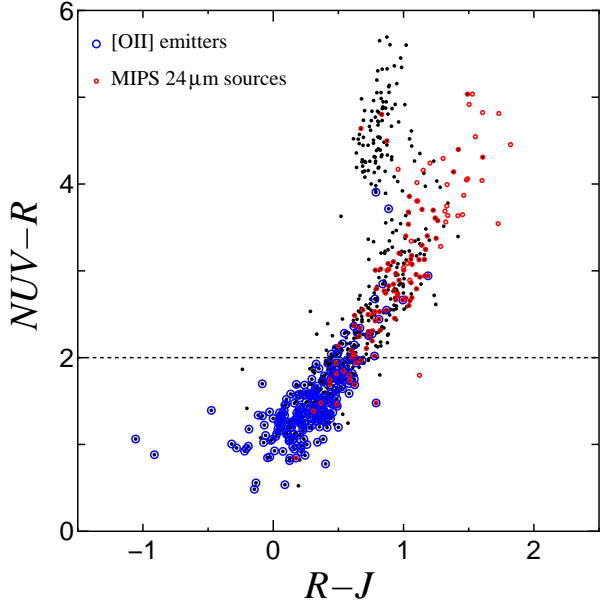} 
\caption{
The rest-frame $NUV - R$ vs. $R - J$ diagram for the Sample B ($M_{U3500} < -19.31$).
Blue circles show \OII\ emitters selected by the NB711-band excess.
Red circles represent bright Spitzer/MIPS 24$\mu$m sources with $f_{\rm 24\mu m} 
\gtrsim $ 150--200 $\mu$Jy 
 at $z_{\rm phot} = $0.901--0.920,  
which include those objects with $M_{U3500} > -19.31$. 
}
\label{fig:twocolor}
\end{center}
\end{figure}

In order to check whether they affect the environmental dependence of the fraction of 
star-forming galaxies, we selected star-forming galaxies by the rest-frame 
$NUV - R$ color estimated from the COSMOS multi-band photometric data instead of 
$EW_0$(\OII). The $EW_0$(\OII) is the ratio of the line luminosity 
$L([{\rm O{\mathsc{\ ii}}}])$ and 
the continuum luminosity at the rest-frame 3727\AA . 
While $L([{\rm O{\mathsc{\ ii}}}])$ mainly depend on 
SFR, the continuum luminosity at 3727\AA\ 
depends on both SFR and stellar mass. Therefore we consider 
that the rest-frame $NUV - R$ color is more suitable for a substitute for 
$EW_0$(\OII) than the simple rest-frame UV luminosity. 
If we ignore the effect of the dust reddening, star-forming galaxies with large $EW_0$(\OII) 
are expected to have blue $NUV - R$ colors. 
Since the rest-frame $NUV-R$ selection  
is not limited to the narrow redshift range of $z_{\rm phot} = $ 0.901--0.920, 
we can estimate the fraction of blue star-forming galaxies and the local galaxy density 
with the same sample within a redshift slice of $\Delta z = 0.046$.
\begin{figure}[t]
\begin{center}
\includegraphics[width=80mm]{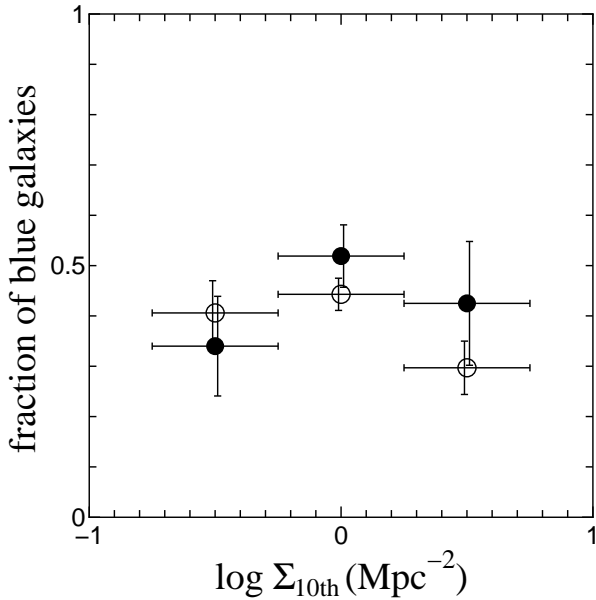} 
\caption{
Fraction of blue galaxies with the rest $NUV - R < 2$ as a function of the local galaxy 
density. Solid circles show the result for galaxies with $M_{U3500} < -19.71$ 
at $0.901 \le z_{\rm phot} \le 0.920$. 
The local density for each sample is measured with galaxies 
within a redshift slice of $z\pm 0.023$ centered on its redshift, as in Figure 
\ref{fig:fraction}. On the other hand, open circles show the case that 
both the fraction of blue galaxies and the local galaxy density are measured from 
the same galaxy sample at $0.91-0.023 \le z_{\rm phot} \le 0.91+0.023$.
}
\label{fig:bluefrac}
\end{center}
\end{figure}

Figure \ref{fig:twocolor} shows the two color diagram of $NUV - R$ vs. $R-J$ for galaxies at  
 $0.901 \le z_{\rm phot} \le 0.920$. It is seen that \OII\ emitters tend to show 
blue $NUV - R$ colors as expected. Since more than 90\% of \OII\ emitters have $NUV - R < 2$, 
we use $NUV - R < 2$ as the selection criterion for blue star-forming galaxies. 
The average fraction of galaxies with $NUV - R < 2$ is 47\% in the Sample A, which 
is slightly larger than that of \OII\ emitters. 
Figure \ref{fig:bluefrac} shows the fraction of galaxies with $NUV - R <2$ in the Sample A 
as a function of the local galaxy density. We first used galaxies at 
$0.901 \le z_{\rm phot} \le 0.920$ as the main sample and measured the local density for 
each sample galaxy by using galaxies within a redshift slice of $z\pm\sigma_z$ centered on 
its redshift, as in the case of \OII\ emitters (solid circles in Figure \ref{fig:bluefrac}). 
Then we estimated both the fraction of blue galaxies and the local galaxy density from 
the same galaxies at $0.91-0.023 \le z_{\rm phot} \le 0.91+0.023$ 
(open circles in the figure). The fractions of galaxies with $NUV - R <2$ in the both cases 
agree within the uncertainty at each density. The fraction does not significantly depend on 
the local density in the both cases, although the fraction could be slightly 
smaller at the highest density bin in the latter case. 
If we use the Sample B instead of the Sample A, we obtain the similar results. 
Therefore we consider that the difference in the redshift range between the main sample and 
the sample used for the density measurement does not significantly affect our results.

\subsection{Comparison with the semi-analytic model}
\label{sec:semiana}

Although more than 30 photometric bands of the COSMOS data set provide the very accurate 
photometric redshift, the error of $\sigma_z = 0.023$ at $z\sim0.9$ corresponds to 
relatively large physical scale of $\sim 30$ Mpc. The effects of the projection over 
the redshift slice of $z\pm\sigma_z$ ($\sim 60$Mpc width in physical scale) and the 
photo-z error could wash out structures or make spurious overdensities. 
\begin{figure*}
\begin{center}
\includegraphics[width=160mm]{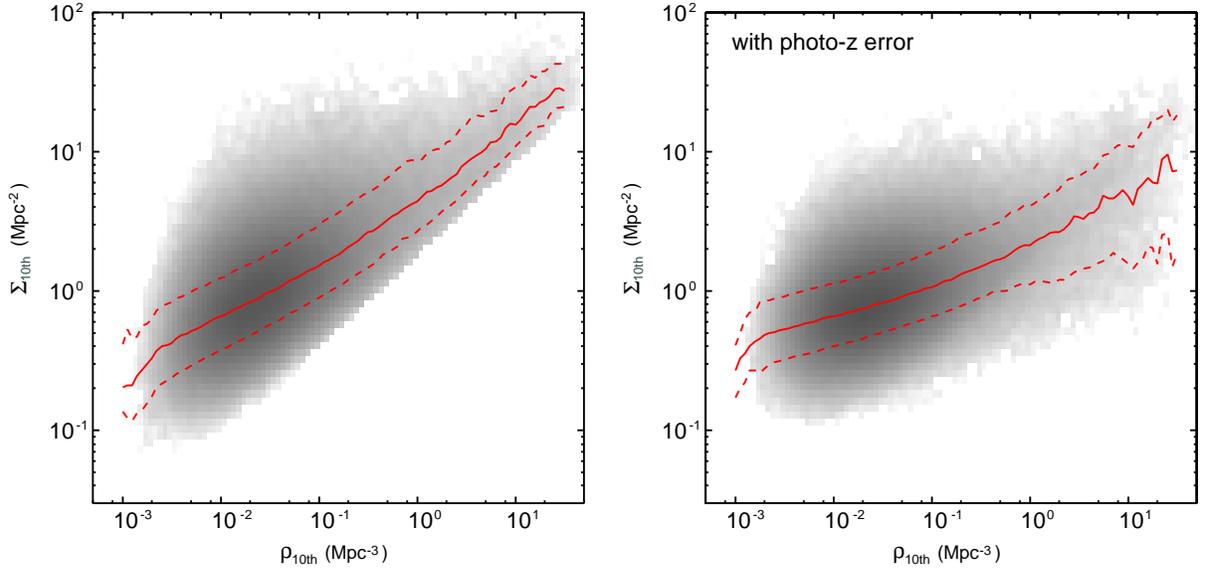} 
\caption{
{\bf left)} Comparison between the 3-dimensional (true) local galaxy density and 
the projected 2-dimensional density for $\sim 550000$ mock galaxies 
with $M_{U} < -19.71$ at $z\sim0.9$ 
in the semi-analytic model by \cite{fon08}. The both densities are estimated by 
the 10th nearest neighbor method. The 2-dimensional projected density is calculated 
from model galaxies within a slice of 60 Mpc in physical scale, which corresponds to 
 $\Delta z = 0.046$ at $z\sim0.9$. 
Solid line shows the median value of the projected density as a function of 
the 3-dimensional density, while dashed lines represent the 16 and 84 percentiles. 
{\bf right)} The same as the left panel but the projected density is calculated from 
model galaxies within a slice of 60 Mpc after galaxies are randomly shifted 
with a offset of the Gaussian distribution with $\sigma = 30$ Mpc 
in order to take account of the photometric redshift error (see text). 
}
\label{fig:sigcheck}
\end{center}
\end{figure*}

In order to check these effects, we use a mock galaxy catalogue from the publicly 
available semi-analytic model by \cite{fon08}. This model is based on the Millennium 
Simulation of the growth of dark matter structure in a $\Lambda$ cold dark matter 
cosmology \citep{spr05} and is a minor revision of the model by \cite{bow06}.
We sampled model galaxies with $M_{U} < -19.71$ and $M_{U} < -19.31$ from 
a snapshot at $z\sim0.9$, and then calculated the local galaxy densities with three 
ways; 1) 3-dimensional density based on the 10th nearest neighbor (i.e., true density), 
2) 2-dimensional projected density based on the 10th nearest neighbor 
which is calculated from samples within a slice of 60 Mpc in physical scale, 
3) the same as 2) but calculated from samples within a slice of 60 Mpc 
after galaxies are randomly shifted along the depth direction of the slice with a offset  
of the Gaussian distribution with $\sigma = 30$ Mpc in order to take the photometric 
redshift error of the observed sample into account. 
We can check the projection effect by comparing 1) and 2), and see the effect of the 
photometric redshift uncertainty from 3). Figure \ref{fig:sigcheck} shows the comparison 
between 1) and 2) (left panel) and that between 1) and 3) (right panel) for the sample with 
$M_{U} < -19.71$.  
It is seen from the left panel that there is a clear correlation with a scatter of 
$\sigma \sim 0.3$ dex between the true density and 2-dimensional projected density.
Even if the random offsets by the photometric redshift error are added (the right panel), 
there remains the correlation between the true and projected densities although 
both high and low density regions seem to be smeared out in some degree 
and the correlation becomes weaker. The results for the sample with $M_{U} < -19.31$ 
are the same except that the both true and projected densities become slightly 
larger simply because of the increase of the number of galaxies in the sample. 
We confirmed that these results do not depend on 
a choice of the direction of the slice in the simulation box.   
\begin{figure*}
\begin{center}
\includegraphics[width=160mm]{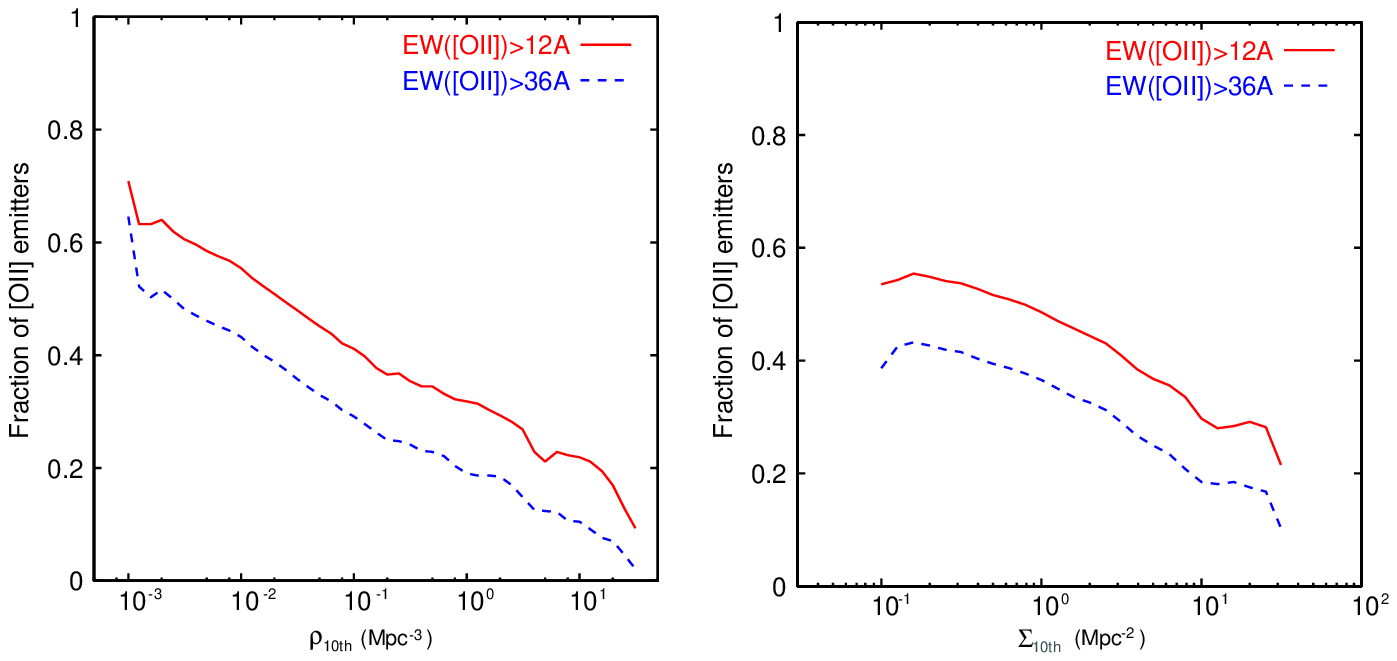} 
\caption{
{\bf left)} Fraction of \OII\ emitters in mock galaxies with $M_{U} < -19.71$ at $z\sim0.9$ 
in the semi-analytic model by \cite{fon08} as a function of the 3-dimensional local 
galaxy density. Solid line shows the fraction of galaxies with $EW_0$(\OII) $> 12$ \AA,
and dashed line shows that of galaxies with $EW_0$(\OII) $> 36$ \AA.
{\bf right)} The same as the left panel but as a function of the 2-dimensional 
projected density with the photometric redshift error of $\sigma_{z} = 0.023$.
}
\label{fig:fracfont}
\end{center}
\end{figure*}

We also checked the environmental dependence of the fraction of \OII\ emitters 
in the simulation by calculating $EW_0$(\OII) of model galaxies from their 
 $L([{\rm O{\footnotesize{II}}}])$ and $M_{U}$ in the mock catalogue.
Figure \ref{fig:fracfont} shows the fraction of galaxies with 
$EW_0$(\OII) $> 12$ \AA\ as a function of the 3-dimensional density (left panel) and 
the 2-dimensional projected density with the photometric redshift error (right panel). 
Since the overall fraction is slightly larger than the observed fraction of \OII\ emitters, 
we also show the fraction of model galaxies with $EW_0$(\OII) $> 36$ \AA\ for comparison. 
It is seen from the left panel that the fraction of star-forming galaxies 
selected by the $EW_0$(\OII) criteria clearly decreases with the local galaxy density in the 
semi-analytic model as previous studies have already reported (e.g., \citealp{elb07}). 
Even if we use the projected density with the photometric redshift error, the fraction of 
\OII\ emitters clearly depends on the density, although the environmental dependence 
of the fraction becomes slightly weaker. This suggests that the effects of the 
projection and the photometric redshift error in our density measurement do not 
smear out the relation between the fraction of \OII\ emitters and local density. 
The observed fraction in Figure \ref{fig:fraction} does not seem to decrease 
with the local density, which is different from the semi-analytic model, 
although the relatively large statistical error in our analysis 
prevents us from completely rejecting the model. 

We note that the above simulation of the fraction of \OII\ emitters with the semi-analytic 
model does not consider the fact that \OII\ emitters are selected by the narrow-band excess 
within a narrower redshift width of $\Delta z = 0.019$. As seen in the previous 
subsection, however, we obtained the same result (no significant environmental 
dependence) when we selected galaxies 
with blue rest-frame $NUV - R$ colors as the star-forming population within the 
same redshift width of $\Delta z = 0.046$ as in the local density measurement. 
Therefore, we consider that it does not affect the comparison between our result 
and the model prediction.

\subsection{Incompleteness and contamination due to the photometric redshift error}
\label{sec:specz}

We here check the incompleteness and contamination due to the photometric redshift 
error in our sample by using spectroscopic redshifts from the zCOSMOS redshift survey 
(\citealp{lil07}; \citealp{lil09}).
Since the NB711 data are used in the photometric redshift measurement \citep{ilb09}, 
the photometric redshifts of \OII\ emitters with the narrow-band excess are expected to be 
more accurate than the other galaxies without the narrow-band excess 
(hereafter, non-\OII\ emitters) in our sample. 
Therefore we investigated the incompleteness and contamination for \OII\ emitters and 
non-\OII\ emitters separately. 

We first examined the spectroscopic redshift distribution for the photo-z selected galaxies 
with $M_{U3500} < -19.71$ (Sample A) in order to estimate the fraction of the contamination 
from outside the redshift range into the photo-z selected sample.
There are 24 \OII\ emitters and 65 non-\OII\ emitters 
with spectroscopic identification in the Sample A, and we can use these galaxies for the 
purpose. 
Out of 24 \OII\ emitters, 19 have spectroscopic redshifts of $0.901 \le z_{\rm spec} \le 0.920$, 
and the other 5 objects have spectroscopic redshifts outside of the redshift range 
(i.e., $z_{\rm spec} < 0.901$ or $z_{\rm spec} > 0.920$). 
If we consider these 5 objects as the contaminants from outside the redshift range, 
the fraction of the contamination from outside the redshift range is $5/24 = 21$\%.
Actually, four of these 5 objects have slightly lower redshifts of 
$z_{\rm spec} =$ 0.893--0.899, for which the \OII\ emission enters into the short-wavelength 
wing of the NB711 filter, while one object is a [O {\footnotesize III}] emitter at 
$z_{\rm spec} = 0.422$. So if we include four objects with $z_{\rm spec} =$ 0.893--0.899 into  
the \OII\ emitter sample, the contamination rate becomes $1/24 = 4$\%. On the other hand, 
39 of 65 non-\OII\ emitters in the Sample A lie within $0.901 \le z_{\rm spec} \le 0.920$, 
and the other 26 objects have spectroscopic redshifts of 
$z_{\rm spec} < 0.901$ or $z_{\rm spec} > 0.920$.
Therefore, the contamination rate for non-\OII\ emitters is $26/65 = 40$\%. 

Next, we checked the photometric redshift distribution of 
galaxies with $z_{\rm spec} =$ 0.901--0.920 and 
$M_{U3500} < -19.71$ to estimate the incompleteness due to the photo-z error for the Sample A.  
From the spectroscopic catalogue, we could use 97 objects with $z_{\rm spec} =$ 0.901--0.920 
and $M_{U3500} < -19.71$, and found that 58 out of 97  
have photometric redshifts of $0.901 \le z_{\rm phot} \le 0.920$.
While these 58 objects are included into our photo-z selected sample,  
the other 39 objects are missed by the photo-z selection. 
Out of 97 objects with $z_{\rm spec} =$ 0.901--0.920, 20 objects shows a significant 
NB711-band excess and satisfy the criteria for \OII\ emitters, and the other 
77 objects are non-\OII\ emitters. Similarly, out of 58 objects 
with $z_{\rm spec} =$ 0.901--0.920 and $z_{\rm phot} =$ 0.901--0.920, 
19 objects satisfy the criteria for \OII\ emitters, and the other 39 objects are 
non-\OII\ emitters. Therefore, for \OII\ emitters, 19 out of 20 galaxies with 
 $z_{\rm spec} =$ 0.901--0.920 are selected to the Sample A, and the completeness 
is $19/20 = 95$\%. On the other hand, 39 out of 77 non-\OII\ emitters with 
$z_{\rm spec} =$ 0.901--0.920 are included into the sample. Therefore 
the completeness for non-\OII\ emitters is $39/77 = 51$\%.

By adopting these contamination and completeness rates, which is estimated 
from the spectroscopic sample, for all the photo-z selected sample, 
we examined the effects of the contamination and incompleteness due to the photo-z error 
on the fraction of \OII\ emitters in the photo-z selected sample. 
Since the contamination and completeness rates for \OII\ emitters are 21\% and 95\%, 
respectively, we select 95\% of the real \OII\ emitters and also pick up  
the contaminants which account for 21\% of the observed number. Therefore, 
the number of \OII\ emitters are expected to be overestimated by $\sim $20\%; 
i.e., $0.95/(1-0.21) \sim 1.20$. 
Similarly, from the contamination rate of 40\% and the completeness of 51\%, 
we can calculate that the observed number of non-\OII\ emitters is underestimated 
by $\sim $15\%; i.e., $0.51/(1-0.40) \sim 0.85$.
Since the observed numbers of \OII\ emitters and non-\OII\ emitters are 95 and 196 respectively, the number of all photo-z selected galaxies 
is expected to be underestimated by $\sim 6$\%; i.e., $(95+196)/(95/1.21  
+ 196/0.85) \sim 0.94$. As a result, the number of \OII\ emitters is overestimated by 
$\sim$20\%, while the number of the all photo-z sample is underestimated by $\sim 6$\%. 
Therefore, the fraction of \OII\ emitters could be 
overestimated by $\sim $ 28\%; i.e., $1.20/0.94 = 1.28$.
If we do not consider the four \OII\ emitters with $z_{\rm spec} =$ 0.893--0.899 as 
the contaminants, the contamination rate for \OII\ emitters 
decreases from 21\% to 4\% as mentioned above, 
and the overestimation of the fraction of \OII\ emitters becomes $\sim$ 5\%. 
Since \cite{ide12} estimated that the fraction of \OII\ emitters at $z\sim1.2$ could be 
overestimated by $\sim 17$\% due to the photo-z error, the fractions of \OII\ emitters 
at $z\sim1.2$ and $z\sim 0.9$ might be similarly overestimated.

\section{DISCUSSION}
\subsection{Comparison with other studies}

We here compare our results in the COSMOS field
 with previous studies of the environmental dependence of the star formation activity 
in galaxies at similar redshifts. 
\cite{elb07} investigated the average SFR of galaxies with $M_{B}<-20$ 
at $0.8 \le z \le 1.2$ as a function of the local galaxy density in the GOODS fields.  
They found that the average SFR increases with the density and seems to peak around 
a density of $\sim $ 2--3 Mpc$^{-2}$. Our results seem to be consistent with their 
results in that the fraction of actively star-forming galaxies does not decreases with 
the local density on average, which is different from the SFR-density relation seen 
in the present universe and from the predictions by the semi-analytic models of 
the galaxy formation. However, the fraction of \OII\ emitters at $z\sim0.9$ 
in the COSMOS field does not significantly 
depend on the density and does not seem to peak around a density of several Mpc$^{-2}$.
In order to examine the environmental dependence of the average SFR in our sample, 
we plot $L([{\rm O{\mathsc{\ ii}}}])$ of \OII\ emitters as a function of the local density 
in Figure \ref{fig:loiisigma}. 
$L([{\rm O{\mathsc{\ ii}}}])$ is estimated from the $ri$ and $NB711$ magnitudes (see 
Appendix for detail). 
It is seen that the $L([{\rm O{\mathsc{\ ii}}}])$ of \OII\ 
emitters is also independent of the density. Therefore we expect that the average 
$L([{\rm O{\mathsc{\ ii}}}])$ also does not significantly depend on the local density.
 \begin{figure}[t]
\begin{center}
\includegraphics[width=80mm]{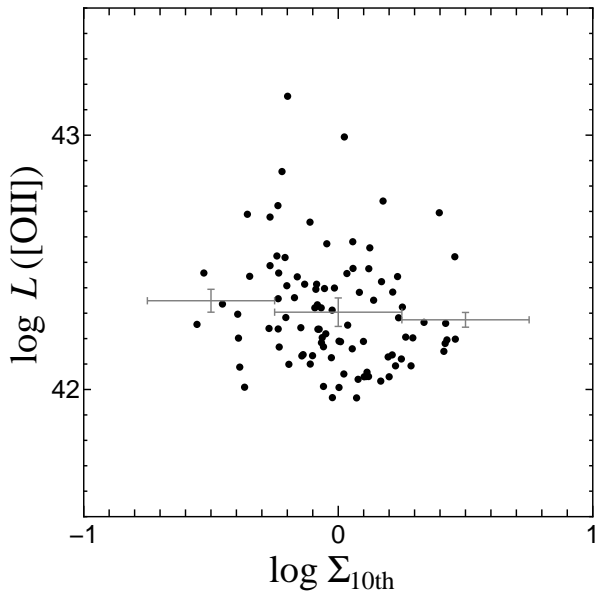} 
\caption{
\OII\ luminosity vs. the local galaxy density 
for \OII\ emitters with $M_{U3500} < -19.71$. 
The \OII\ luminosity is corrected for the dust extinction under the assumption of 
$A_{\rm H\alpha}$ = 1 mag for all \OII\ emitters (see Appendix). 
}
\label{fig:loiisigma}
\end{center}
\end{figure}

\begin{figure}[t]
\begin{center}
\includegraphics[width=80mm]{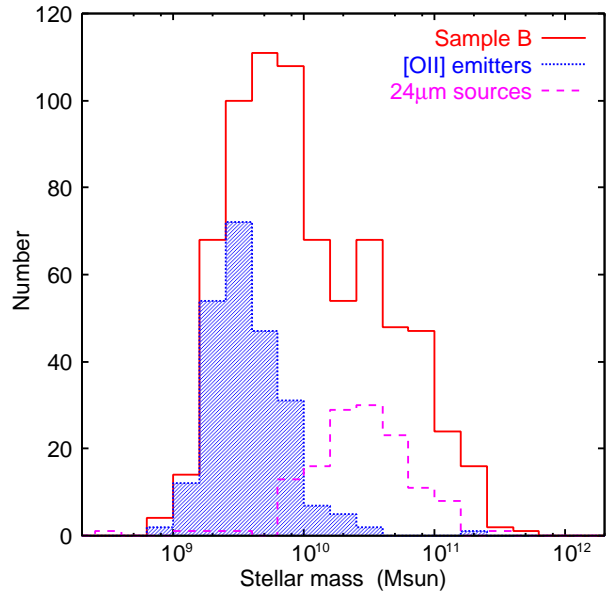} 
\caption{Distribution of the stellar mass of galaxies for the Sample B ($M_{U3500} < -19.31$).
The solid line shows all galaxies in the Sample B, 
while the shaded dotted histogram shows \OII\ emitters with $M_{U3500} < -19.31$.
The dashed histogram represents the bright MIPS 24$\mu$m sources with $f_{\rm 24\mu m} 
\gtrsim $ 150--200 $\mu$Jy. 
}
\label{fig:mstar}
\end{center}
\end{figure}

A possible origin of the different behaviors in the SFR-density relation  
 between \cite{elb07} and this study is 
the effect of the dust extinction. We selected star-forming galaxies with the 
\OII\ emitter selection, while \cite{elb07} estimated SFRs of galaxies from the Spitzer/MIPS 
24 $\mu$m fluxes. Although we correct $L([{\rm O{\mathsc{\ ii}}}])$ for the dust extinction 
assuming $A_{\rm H\alpha}$ = 1 mag, the \OII\ emitter selection itself could 
miss dusty star-forming galaxies. In order to check this, we cross-matched the 24$\mu$m 
source catalogue from the S-COSMOS survey \citep{san07} to galaxies at  
$0.901 \le z_{\rm phot} \le 0.920$ in our sample. The flux limit of the 24$\mu$m catalogue 
is $\sim$ 150--200 $\mu$Jy. We plot the 24$\mu$m sources as red circles 
in Figure \ref{fig:twocolor}. These bright 24$\mu$m sources tend to show red rest-frame 
$NUV - R$ and $R-J$ colors, and the overlap between these 24$\mu$m sources and \OII\ emitters 
is relatively small. Since star-forming galaxies are expected to distribute over a 
diagonal region from ($R - J \sim 0$, $NUV - R \sim 1$) to ($R - J \sim 1.5$, 
$NUV - R \sim 4.5$) (e.g., \citealp{bun10}), the \OII\ selection seems to miss star-forming 
galaxies with relatively red $NUV - R$ colors such as the bright 24$\mu$m sources. 
We also compare the stellar mass of \OII\ emitters with that of these  
bright 24$\mu$m sources in Figure \ref{fig:mstar}.
The stellar mass of each galaxy is estimated by fitting the 
multi-band photometric data from UV to MIR wavelength with the population synthesis model 
by \cite{bru03} \citep{ilb10}. \cite{cha03}'s IMF is assumed. Figure \ref{fig:mstar} 
shows that most of \OII\ emitters have relatively low 
stellar mass of $M_{\rm star}\lesssim 10^{10}M_{\odot}$, 
while the all photo-z selected galaxies distribute over $10^{9}M_{\odot} \lesssim M_{\rm star} 
\lesssim 10^{11}M_{\odot}$. On the other hand, the bright 24$\mu$m sources show systematically 
larger stellar mass of $\sim 10^{10}$--$10^{11}M_{\odot}$.
The 24$\mu$m-selected star-forming galaxies in \cite{elb07} also distribute over
 $10^{9}M_{\odot} \lesssim M_{\rm star} \lesssim 10^{11}M_{\odot}$.  
Therefore the \OII\ emitter selection seems to miss massive (dusty) star-forming population. 
Since star-forming galaxies with $10^{10}$--$10^{11}M_{\odot}$ have a large contribution 
to the cosmic SFR density at $z\sim1$ (e.g., \citealp{kaj10}), these massive star-forming 
galaxies could significantly contribute the average SFR in each environment.
 The contribution from these massive galaxies might cause the peak of the average SFR 
around a local density of $\sim $ 2--3 Mpc$^{-2}$ seen in \cite{elb07}. 
However, we note that the fraction of 
star-forming galaxies in our sample does not significantly depend on the density, 
even if we include the bright 24$\mu$m sources into the star-forming population.
\begin{figure*}[t]
\begin{center}
\includegraphics[width=125mm]{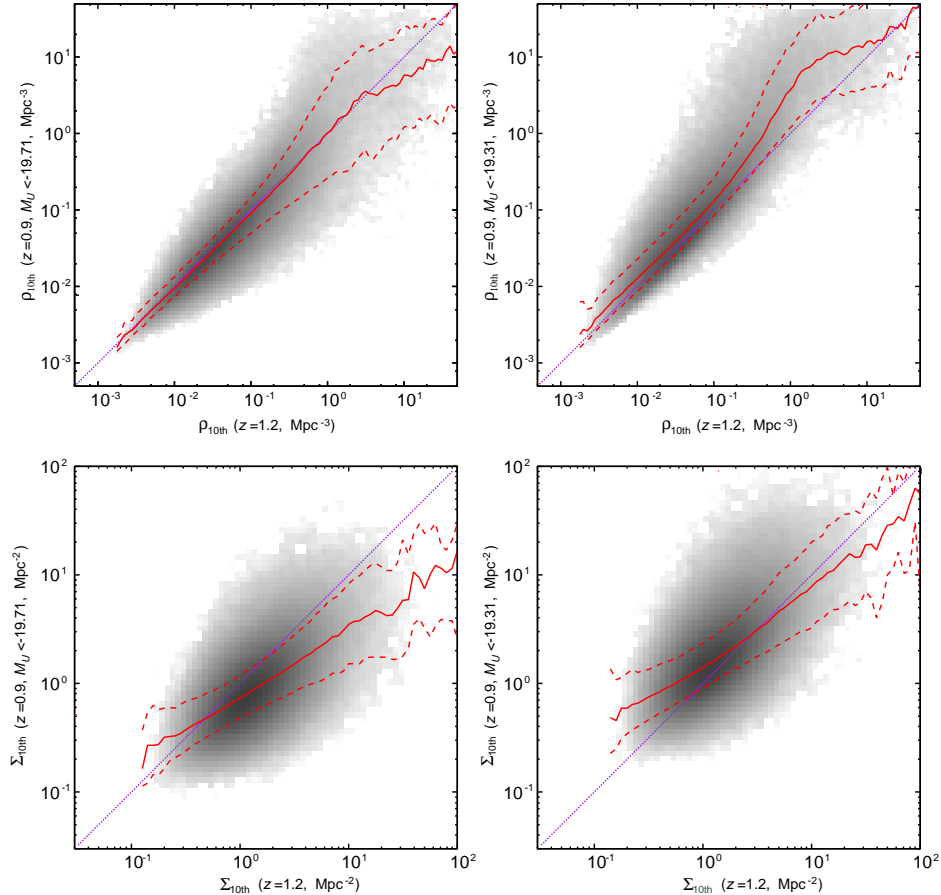} 
\caption{
Evolution of the local galaxy density between $z\sim1.2$ and $z\sim0.9$ 
for mock galaxies in the semi-analytic model by \cite{fon08}.
Top panels show comparisons of the 3-dimensional local densities at the 
different redshifts for the same model galaxies, while bottom panels show 
those of the 2-dimensional projected densities with the photometric redshift 
error. 
The local density at $z\sim1.2$ is estimated from galaxies with $M_{U} < -19.71$.
Left panels show the case that 
the local density at $z\sim0.9$ is estimated from galaxies with $M_{U} < -19.71$,  
while right panels show the results for the density at $z\sim0.9$ measured from 
those with $M_{U} < -19.31$. 
Solid line shows the median value of the density at $z\sim0.9$ as a function of 
the density at $z\sim1.2$, while dashed lines represent the 16 and 84 percentiles. 
}
\label{fig:sigmaev}
\end{center}
\end{figure*}

Another possible origin of the different results is the different scales of the local density 
measurement. \cite{elb07} measured the density with a box of 1.5 Mpc $\times$ 1.5 Mpc $\times$ 
40 Mpc in comoving scale, while we used the 10th nearest neighbor method with galaxies
 within a comoving depth of 118 Mpc. 
In the 10th nearest neighbor method, a radius used in the density measurement depends on 
the local density itself. For example, the local density of  
$\Sigma_{\rm 10th} = $ 1 Mpc$^{-2}$ corresponds to a radius of $r_{\rm 10th} \sim 3.6$ Mpc 
in comoving scale ($\sim 1.9$ Mpc in physical scale). 
When the local density is 10 Mpc$^{-2}$, a radius becomes 1.1 Mpc in comoving scale. 
Since we typically investigate galaxies with 
$\Sigma_{\rm 10th} \sim $ 1--10 Mpc$^{-2}$ (Figure \ref{fig:fraction}), 
the local density in our analysis is measured with a radius of 1.1--3.6 Mpc, 
which corresponds to a diameter of 2.2--7.2 Mpc.  
This scale is larger than that in \cite{elb07} (1.5 Mpc), especially for low-density 
environments. The depth of 118 Mpc used in our density measurement is also much 
larger than 40 Mpc used in \cite{elb07}. 
For local galaxies, \cite{bla06} pointed out that the environment dependence of the 
star formation activity in galaxies could vary with the scale used for the density measurement.
\cite{elb07} also discussed that the peak of the average SFR around 
a density of $\sim $ 3 Mpc$^{-2}$ seen in their result 
could be caused by active star formation in galaxies 
during group formation at the scale of $\sim 1$ Mpc. 
The density measurement in a relatively large scale 
might smear out such small-scale environmental effect  
in some degree. 
\begin{figure*}[t]
\begin{center}
\includegraphics[width=130mm]{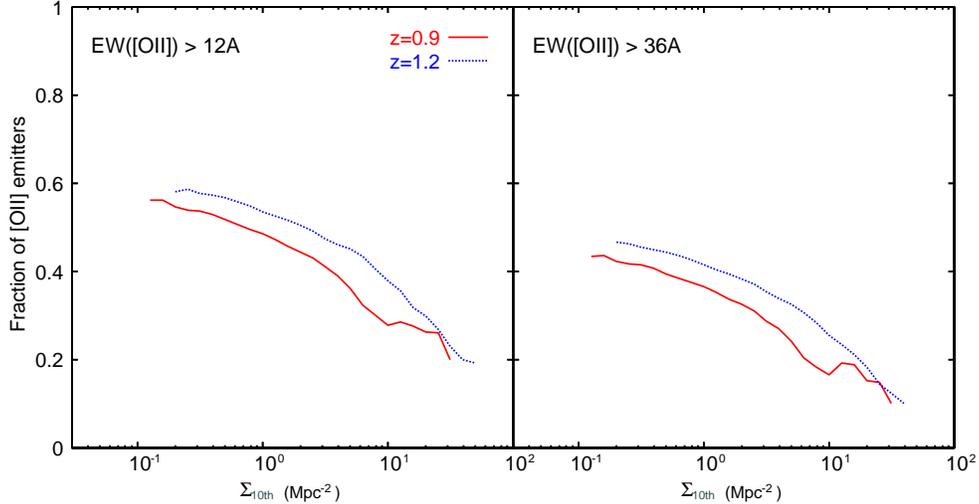} 
\caption{
Evolution of the fraction of \OII\ emitters between $z\sim1.2$ and $z\sim0.9$ 
for mock galaxies with $M_{U} < -19.71$ in the semi-analytic model as a function of 
the 2-dimensional projected density with the photometric redshift error at each 
redshift. Solid line shows the fraction at $z\sim0.9$, while dashed line represents 
that at $z\sim1.2$. Left panel shows the fraction of model galaxies with 
$EW_0$(\OII) $> 12$ \AA, while right panel shows 
the fraction of those with $EW_0$(\OII) $> 36$ \AA. 
}
\label{fig:fracevfont}
\end{center}
\end{figure*}

On the other hand, \cite{pog08} studied the star formation activity of galaxies as 
a function of the local density mainly in clusters and groups at $z=$0.4--0.8.
They found that the fraction of \OII\ emitters in groups or outskirts of clusters 
is similar with that in field environments at the same redshift, while the fraction 
decreases toward the highest-density region at the cores of clusters.
The weak or no environmental dependence of the fraction of \OII\ emitters 
at the relatively low-density environment ($\Sigma_{\rm c} < $ 15--40  ${\rm Mpc^{-2}}$) 
might be consistent with our results,  
although the fraction of $\sim 70$\% in \cite{pog08} is much higher 
than that in this study because 
they selected objects with $EW_0$(\OII) $> 3$ \AA\ as \OII\ emitters by using 
the spectroscopic data. 
They also pointed out that the average SFR of \OII\ emitters 
has a peak at $\Sigma_{\rm c} \sim $ 15--40 ${\rm Mpc^{-2}}$, while 
we found no significant environmental dependence of the \OII\ luminosity at the 
lower density. 
\cite{pat11} similarly studied SFRs of galaxies at $0.6<z<0.9$ as a function of local density 
including the group and cluster environments, and 
 found that both the fraction of star-forming galaxies and the average 
SFR of star-forming galaxies decrease at densities much higher than the field environments.  
But we note that their sample is limited to relatively massive galaxies with 
$M_{\rm star}>10^{10.25}M_{\odot}$, while \OII\ emitters in this study tend to have 
smaller mass of $M_{\rm star}<10^{10}M_{\odot}$ as seen in Figure \ref{fig:mstar}. 

Recently, \cite{sob11} investigated the fraction of H$\alpha$ emitters in 
galaxies with $K<23$ at $z \sim 0.84$ in the COSMOS and UKIDSS UDS fields 
(total $\sim 1.3$ deg$^2$),  using a narrow-band observations at 1.211 $\mu$m. 
They also used the 10th nearest neighbor method to measure the local density and studied 
the environmental dependence of the fraction of star-forming galaxies.
They found that the fraction of H$\alpha$ emitters with 
$EW_{\rm obs}({\rm H}\alpha) > 50$ \AA~ is $\sim 30\%$ almost independent of the 
local density in relatively low-density environments of their 
$\Sigma_{\rm c} < 10 \; {\rm Mpc^{-2}}$. On the other hand, the fraction rapidly 
decreases with increasing the local density in higher-density environments.
If the environments we investigated are mainly field environments, 
our result at $z\sim0.9$ is consistent with that in \cite{sob11}. 
In this study, we also found that the evolution of the fraction of \OII\ emitters 
between $z\sim1.2$ and $z\sim0.9$ is also independent of the galaxy environment 
in the range of the environment we investigated. 
\cite{sob11} suggested that the star formation in the field environment 
is dominated by normal (non-interacting) galaxies, while 
the star formation in the rich group and cluster environment 
is dominated by the potential mergers (e.g., \citealp{ide12}; \citealp{hay10}). 
If this is the case, the galaxy interactions might not play a important role 
in the star formation activity of most field galaxies. 
As a result, the evolution of star formation activity in these galaxies between 
$z\sim1.2$ and $z\sim0.9$ does not depend on the galaxy environment, 
but might simply be related with physical properties of each galaxy such as gas mass fraction.

\begin{figure*}
\begin{center}
\includegraphics[width=75mm]{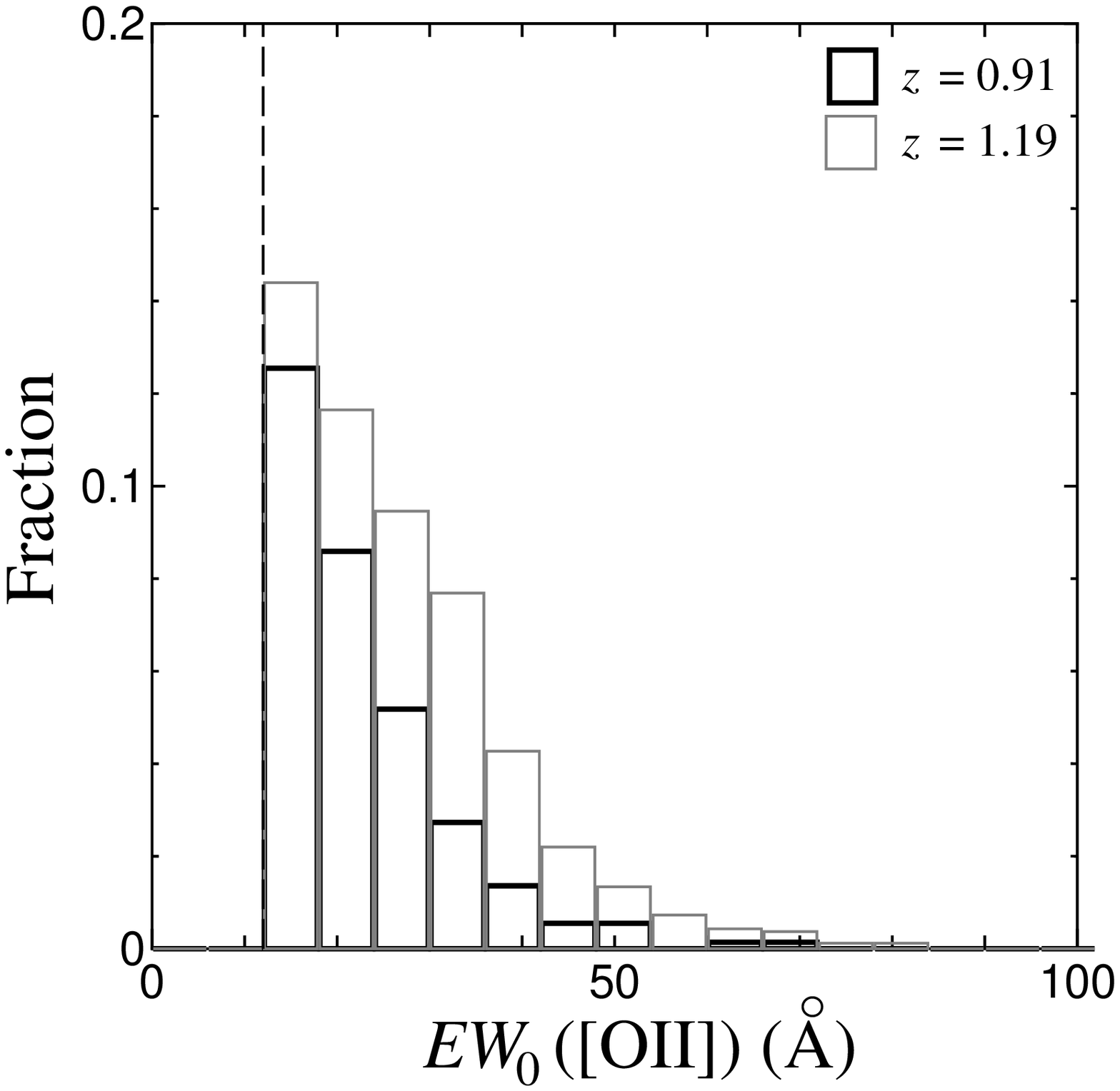} 
\includegraphics[width=75mm]{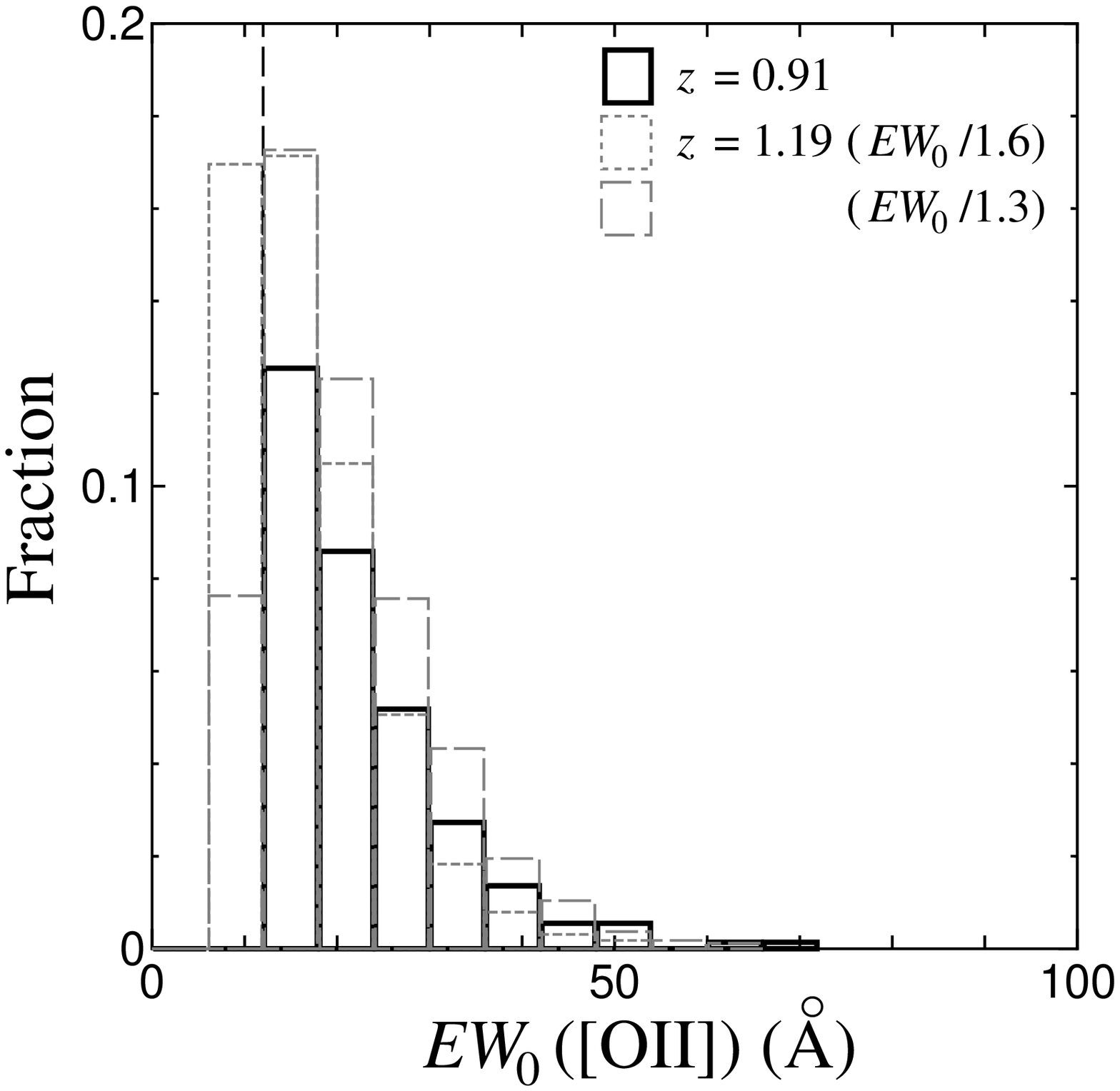} 
\caption{
{\bf left)} Comparison of the normalized 
differential distributions of $EW_0$(\OII)  
between $z\sim 0.9$ and $z\sim1.2$.
black and grey histograms show that for galaxies at $z\sim0.9$ and 
$z\sim1.2$, respectively.
Vertical dashed line shows the equivalent width limit of 
$EW_0$(\OII) $= 12$ \AA~ for our \OII\ emitter selection.
Note that only galaxies classified as \OII\ emitters are plotted, while 
the calculation of the normalization constant takes account of 
all galaxies including objects which do not satisfy the selection criteria 
for the \OII\ emitter. 
{\bf right)} 
The same as the left panel, but the observed $EW_0$(\OII) for \OII\ emitters 
at $z \sim 1.2$ is divided by a factor of 1.3 and 1.6 (short- and long-dashed histograms). 
A decrease of a factor of 1.3--1.6 in $EW_0$(\OII) from $z\sim1.2$ to $z\sim0.9$ 
is expected from the simple model where SFRs of star-forming galaxies decreases 
by a factor of 2 in the redshift interval (see text). 
}
\label{fig:EWdf}
\end{center}
\end{figure*}
\begin{figure}
\begin{center}
\includegraphics[width=80mm]{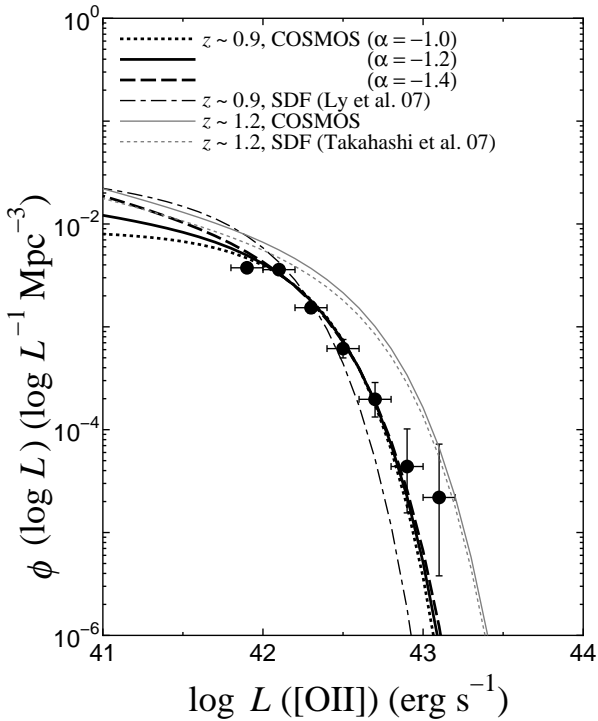} 
\caption{
The \OII\ luminosity function for \OII\ emitters at $z\sim 0.9$ and $z\sim1.2$.
The solid circles show the result for our \OII\ emitter sample at $z\sim0.9$ 
in the COSMOS field. 
The thick dotted, solid, and dashed lines 
represent the best-fit Schechter functions 
for $\alpha=-1.0$, $-1.2$, and $-1.4$, respectively. 
Thin dashed-dotted line shows \OII\ LF for \OII\ emitters 
at $z \sim 0.9$ in the Subaru Deep Field from \cite{ly07}.  
Thin solid line represents the result at $z\sim1.2$ in the 
COSMOS field updated from \cite{tak07}  
with the newest version of the COSMOS photometric redshift catalog.
Thin dotted lines show \OII\ LF at $z \sim 1.2$ in 
the Subaru Deep Field from \cite{tak07}. 
}
\label{fig:LF}
\end{center}
\end{figure}

\subsection{Evolution of the fraction of \OII\ emitters between $z\sim1.2$ and $z\sim0.9$}

In this section, we discuss the strength of the evolution in the fraction of \OII\ emitters 
from $\sim 60\%$ at  $z \sim 1.2$ to $\sim 30\%$ at  $z \sim 0.9$. 
At first, we examined how 
the local density for each galaxy evolves between $z \sim 1.2$ and $z\sim0.9$ 
by using the same semi-analytic model as in Section 
\ref{sec:semiana},  
since the strong evolution of the density could make the comparison between 
the different redshifts more complicated (e.g., \citealp{pog10}). 
Figure \ref{fig:sigmaev} shows the comparisons of the local density between the redshifts 
for the same model galaxies for the Sample A and Sample B.  
For model galaxies at $z\sim1.2$, we similarly calculated the 3-dimensional density and 
the 2-dimensional projected density including 
 the random offsets due to the photo-z error at $z\sim1.2$. 
Note that the average number densities of galaxies at $z\sim0.9$ 
are different between the both samples by a factor of $\sim 2$. 
It is seen from the upper panels of the figure 
that the evolution of the true 3-dimensional density is relatively small 
(a factor of $\lesssim 2$) for most model galaxies except for the high density region. 
The projected density also mildly evolves at 0.3 Mpc$^{-2} \le \Sigma_{\rm 10th} \le $ 
10 Mpc$^{-2}$ especially for the Sample B. 
By combining this with the lack of the environmental dependence at the both redshifts, 
we infer that the change in the local density probably does not strongly affect 
the evolution of the fraction of \OII\ emitters in the range of the density we investigated.
Therefore we can fairly compare the fractions of \OII\ emitters 
at the same range of the local density between $z\sim1.2$ and $z\sim0.9$.

In Figure \ref{fig:fracevfont}, 
we plot the evolution of the fraction of \OII\ emitters between 
$z\sim0.9$ and $z\sim1.2$ predicted by the semi-analytic model.  
The strength of the evolution in the model does not seem to be affected by the criterion of 
$EW_0$(\OII), and it could be slightly large at $\Sigma_{\rm 10th} \sim $ 3--10 Mpc$^{-2}$.  
The observed strength of the evolution seen in Figure \ref{fig:fraction} 
is larger than that in the model.  

We next consider the evolution of the $EW_0$(\OII) distribution. 
Figure \ref{fig:EWdf} shows a comparison of the observed $EW_0$(\OII) distribution between 
the $z\sim0.9$ and $z\sim1.2$ samples. 
The $EW_0$(\OII) for each \OII\ emitter is calculated from the narrow-band excess 
(NB711 $-$ $ri'$). 
In Figure \ref{fig:EWdf}, the fraction of galaxies with $EW_0$(\OII) $>$ 12\AA\ at 
$z\sim0.9$ is lower than that at $z\sim1.2$ as expected from Figure \ref{fig:fraction}.  
If we simply assume the $EW_0$(\OII) of all star-forming galaxies decreases by the 
same factor from $z\sim1.2$ and $z\sim0.9$, the observed $EW_0$(\OII) distributions 
suggest that the $EW_0$(\OII) needs to decrease by a factor of 
$1.75^{+0.55}_{-0.35}$ in this redshift interval 
in order to reproduce the fraction of \OII\ emitters with $EW_0$(\OII) $\geq 12$ \AA~ 
at $z\sim0.9$ ($\sim 0.3 \pm 0.1$). 

We here try to interpret the $EW_0$(\OII) evolution as the evolution of star 
formation activity in galaxies. Figure \ref{fig:LF} compares \OII\ LFs between 
$z\sim 0.9$ and $z\sim 1.2$ in the COSMOS field and the Subaru Deep Field (SDF).  
The \OII\ LFs in the COSMOS field are derived from our samples at 
$z\sim 0.9$ and $z\sim 1.2$ (Appendix A), and those in the SDF are the results by 
\cite{ly07}. It is seen that the characteristic luminosity decreases 
by a factor of $\sim 2$ from $\log L_* \sim 42.5$ at $z\sim1.2$ to $\log L_* \sim 42.2$ at 
$z\sim 0.9$ in the both fields, while the normalization shows no significant evolution. 
If we assume that there is no strong evolution in the average metallicity of galaxies 
between $z\sim1.2$ and $z\sim0.9$, this luminosity evolution reflects the decrease 
of the star formation activity in galaxies. Many previous studies of the 
evolution of the cosmic SFR density also suggest that 
the star formation activity in the universe decreases by a factor of $\sim 2$ 
from $z\sim 1.2$ to $z\sim0.9$ on average (e.g., \citealp{hop06}; \citealp{sob12}).
Then we calculate the evolution in the $EW_0$(\OII) for the case that the star 
formation rate decreases by a factor of 2 in the redshift interval. 
The \OII\ line luminosity is simply expected to decreases by the same factor 
as the SFR, if we ignore the metallicity/dust extinction evolution. 
If we assume exponentially decaying star formation histories (i.e., 
SFR $\propto \exp(-{\rm age}/\tau)$) between $z\sim 1.2$ and $z\sim0.9$ for simplicity, 
the evolution of factor of $\sim 2$ in the SFR  
which is expected from the \OII\ luminosity function 
at $z\sim1.2$ and $z\sim0.9$ 
 corresponds to the star formation timescale of $\tau = $1.2--1.8 Gyr. 
In this case, the continuum luminosity at the rest-frame 3727 \AA~ is expected 
to decrease by a factor of $\sim$ 1.5.
This is also consistent with the evolution of the characteristic magnitude 
of the rest-frame 3500 \AA~ LF (0.4 mag) mentioned in Section 2.1.
Therefore, we can expect that the $EW_0$(\OII) decreases 
by a factor of 1.3--1.6 between $z\sim1.2$ and $z\sim0.9$, 
taking account of the possible effect of 
the Balmer break on the measurement of the narrow-band excess (NB711 $-$ $ri'$).
This is roughly consistent with the evolution of the $EW_0$(\OII) expected from 
the evolution in the fraction of \OII\ emitters ($1.75^{+0.55}_{-0.35}$).  
In the right panel of Figure \ref{fig:EWdf}, we compare the observed $EW_0$(\OII) distribution 
at $z\sim0.9$ with the simulated distributions which are calculated by dividing 
the $EW_0$(\OII) at $z\sim1.2$ by a factor of 1.3 and 1.6. 
The agreement of the $EW_0$(\OII) distribution is relatively good especially in  
the case that the $EW_0$(\OII) decreases by a factor of 1.6, although the observed 
$EW_0$(\OII) at $z\sim0.9$ could be slightly lower than those predicted from the 
evolution of the star formation activity. 
Thus, the decrease of the fraction of \OII\ emitters from $z\sim1.2$ to $z\sim0.9$ 
seems to be explained mainly by the decrease of the overall star formation activity in 
the universe, while the additional offset seen in the right panel of Figure \ref{fig:EWdf} 
might be explained by the evolution of the gas metallicity and/or dust extinction in 
star-forming galaxies.

\section{Summary}

We investigated the fraction of \OII\ emitters in the photo-z selected galaxies 
at $z\sim0.9$ as a function of the local galaxy density in the COSMOS field. 
\OII\ emitters are selected by the narrow-band excess technique with the NB711-band 
data taken with Subaru/Suprime-Cam. We used the magnitude limits and selection criteria 
for \OII\ emitters which are consistent with our previous study at $z\sim1.2$ 
in order to make a direct comparison between $z\sim0.9$ and $z\sim1.2$. 
Our final sample consists of 614 (291) photo-z selected galaxies with 
$M_{U3500} < -19.31$ ($M_{U3500} < -19.71$) at $z=0.901$ -- 0.920, 
which includes 195 (95) \OII\ emitters. Our main results are as follows.

\begin{itemize}

\item
The fraction of \OII\ emitters at $z\sim0.9$ is nearly constant ($\sim 0.3$) 
at $0.3 \; {\rm Mpc^{-2}} < \Sigma_{\rm 10th} < 10 \; {\rm Mpc^{-2}}$. 
The flat distribution holds, 
even if we use the magnitude limit for which the luminosity 
evolution of galaxies is taken into account. 
No significant environmental dependence is similar with the result in 
our previous study at $z\sim1.2$.  

\item 
The fraction of \OII\ emitters decreases from $\sim 0.6$ at $z\sim1.2$  to 
$\sim 0.3$ at $z\sim0.9$ in all environment we investigated. 

\item 
Instead of \OII\ emitters, we used galaxies with blue rest-frame colors 
of $NUV-R<2$ at $0.91-0.023 \le z_{\rm phot} \le 0.91+0.023$ as the star-forming population 
in order to measure both the fraction of star-forming 
galaxies and local galaxy density from the same sample. 
The fraction of blue galaxies with $NUV-R<2$ also does not significantly depend on the local 
density.

\item 
We checked the effects of the projection over the redshift slice and the photometric 
redshift error on our density measurement, using the semi-analytic model by \cite{fon08}. 
Although these effects seem to smear out very high and low density regions in some degree, 
we confirmed that the fraction of \OII\ emitters clearly depends on the projected density 
in the simulation, which is different from the observed results. 

\item 
Most of \OII\ emitters have relatively small stellar mass of $M_{\rm star} < 10^{10}M_{\odot}$, 
and the overlap between \OII\ emitters and bright 24$\mu$m sources with $f_{\rm 24\mu m} 
\gtrsim $ 150--200 $\mu$Jy is relatively small. The \OII\ emitter selection seems to miss 
massive dusty star-forming galaxies. Such a selection bias might cause the different
behaviors in the SFR-density relation among studies with the different SFR indicators.

\item 
If we simply assume SFRs of star-forming galaxies decrease by a factor of 2 from 
$z\sim1.2$ to $z\sim0.9$, which is expected from the evolution of 
the \OII\ luminosity function and cosmic SFR density in the redshift interval, 
the expected evolution of the $EW_0$(\OII) is roughly consistent with 
the observed $EW_0$(\OII) distributions at $z\sim1.2$ and $z\sim0.9$.
Therefore the decrease of the fraction of \OII\ emitters from $z\sim1.2$ to $z\sim0.9$ 
seems to be explained mainly by the decrease of the overall star formation activity in 
the universe.

\end{itemize}

\vspace{1pc}
We would like to thank the referee for invaluable suggestions and comments.
The HST COSMOS Treasury program was supported through NASA grant
HST-GO-09822. We greatly acknowledge the contributions of the entire
COSMOS collaboration consisting of more than 70 scientists. 
This work was financially supported in part by the Japan Society for
the Promotion of Science (Nos. 17253001, 19340046, 23244031, 23654068, and 23740152).

\appendix
\section{Luminosity Function of [O {\sc ii}] Emitters at $z \sim$ 1.2
in the COSMOS Field}

In this Appendix, we describe the derivation of the [O {\sc ii}] luminosity function
at $z \sim 0.9$ in the COSMOS field.
To derive the total \OII\ flux, we have used the total flux density of 
$r^{\prime }$, $i^{\prime }$ (or $i^{*}$ ), and NB711. 
The flux of \OII\ emission line is given by 
\begin{equation}
f_{[{\rm O{\mathsc{\ ii}}}]}=\Delta {\rm NB}\frac{f_{\rm NB}-f_{ri}}{1-0.68(\Delta{\rm NB}/\Delta i^\prime)}.
\end{equation}
where $f_{\rm NB}$ is the total flux density of NB711, 
$f_{ri}$ is the $ri$ continuum flux density, $\Delta$NB and
$\Delta i^\prime$ are the effective bandwidth of the NB711 and $i^\prime$
filters, respectively: $\Delta$NB711 = 72.5 \AA~and $\Delta i^\prime$ =1489.4 \AA.
Since the flux of the \OII\ emission line is affected by the dust obscuration,
it is necessary to correct the extinction effect.
Here, we apply a constant extinction of $A_{[{\rm O{\mathsc{\ ii}}}]}$ = 1.87 mag,
which corresponds to $A_{\rm H\alpha}$ = 1 mag, following previous studies 
(\citealp{hop04}; \citealp{tak07}; \citealp{sob12}).
We also apply the filter transmission effect since the 
actual NB711 filter transmission is not rectangular. 
We adopt a factor of 1.24 following \cite{shi09}.

Then the \OII\ flux is given by 
\begin{equation}
f_{\rm cor}([{\rm O{\mathsc{\ ii}}}]) =f_{[{\rm O{\mathsc{\ ii}}}]}\times 10^{0.4A_{[{\rm O{\mathsc{\ ii}}}]}}\times1.24 \; ,
\end{equation}
and the \OII\ luminosity is estimated by 
\begin{equation}
L([{\rm O{\mathsc{\ ii}}}])=4\pi d_{\rm L}^2 f_{\rm cor}([{\rm O{\mathsc{\ ii}}}])
\end{equation}
where $d_{\rm L}$ is the luminosity distance: $d_{\rm L}$ = 5883 Mpc. 

The \OII\ luminosity function (LF) is constructed by the following formula,
\begin{equation}
\Phi\left(\log L_i\right)=\frac{1}{\Delta\log L}\sum_j\frac{1}{V_j},
\end{equation}
with $|\log L_j-\log L_i|<\frac{1}{2}\Delta\log L$, where $\Delta\log L$ 
is the logarithmic bin size and $V_{j}$ is 
the volume covered by the filter. Here we use 
$\Delta \mathrm{log}\,L=0.2$, and $V_j=2.28\times 10^{5}\ \mathrm{Mpc}\,^{3}$ .
We show the \OII\ LF in Figure \ref{fig:LF}. 

We fit the \OII\ LF with the Schechter function \citep{sch76}, 
\begin{equation}
\Phi(L){\rm d}L=\frac{\phi_\star}{L_\star}\left(\frac{L}{L_\star}\right)^\alpha 
\exp\left(-\frac{L}{L_\star}\right){\rm d}L \; ,
\end{equation}
by the STY method \citep{san79}. 
Before fitting the \OII\ LF, we estimate the lower and upper limiting 
luminosities  ($L_{\rm low}$ and $L_{\rm up}$) that evaluate whether the 
sample is  complete or not. 
Using the observed limiting magnitudes of $ri$ and NB711, 
we obtain $\log L_{\rm low}=41.83\ \rm erg\ s^{-1}$. 
On the other hand, 
the saturation magnitude of $r^\prime$ gives the upper limiting 
luminosity of $\log L_{\rm up}=43.65\ \rm erg\ s^{-1}$.
Since it is difficult to estimate the power index $\alpha$ accurately
because of incompleteness at the faint end, we show our results for the
following three cases, $\alpha = -1.0, -1.2$, and $-1.4$
in Table \ref{tb:parameters}. 
\begin{table}[t]
\begin{center}
\caption{Best-fit Schechter parameters for the \OII\ luminosity function at $z\sim0.9$ in the COSMOS field.}
\label{tb:parameters}

\begin{tabular}{ccc}
\hline\hline
$\alpha$ & $\log L_*$ & $\log\phi_*$ \\
\hline
-1.00 & $42.11 \pm 0.03$ & $-2.43 \pm 0.05$ \\
-1.20 & $42.16 \pm 0.04$ & $-2.48 \pm 0.06$ \\
-1.40 & $42.21 \pm 0.04$ & $-2.55^{+0.07}_{-0.06}$ \\
\hline
\end{tabular}
\end{center}
\end{table}

{}


\begin{thebibliography}{}
\bibitem[Balogh et al.(2004)]{bal04} Balogh, M., Eke, V., 
Miller, C., et al.\ 2004, \mnras, 348, 1355 
\bibitem[Bamford et al.(2009)]{bam09} Bamford, S.~P., Nichol, 
R.~C., Baldry, I.~K., et al.\ 2009, \mnras, 393, 1324
\bibitem[Blanton et al.(2006)]{bla06} Blanton, M.~R., 
Eisenstein, D., Hogg, D.~W., \& Zehavi, I.\ 2006, \apj, 645, 977 
\bibitem[Bower et al.(2006)]{bow06} Bower, R.~G., Benson, 
A.~J., Malbon, R., et al.\ 2006, \mnras, 370, 645 
\bibitem[Bruzual 
\& Charlot(2003)]{bru03} Bruzual, G., \& Charlot, S.\ 2003, \mnras, 344, 1000 
\bibitem[Bundy et al.(2010)]{bun10} Bundy, K., Scarlata, C., 
Carollo, C.~M., et al.\ 2010, \apj, 719, 1969 
\bibitem[Capak et al.(2007)]{cap07} Capak, P., Aussel, H., 
Ajiki, M., et al.\ 2007, \apjs, 172, 99 
\bibitem[Chabrier(2003)]{cha03} Chabrier, G.\ 2003, \pasp, 
115, 763 
\bibitem[Cooper et al.(2008)]{coo08} Cooper, M.~C., Newman, 
J.~A., Weiner, B.~J., et al.\ 2008, \mnras, 383, 1058 
\bibitem[Dressler(1980)]{dre80} Dressler, A.\ 1980, \apj, 
236, 351
\bibitem[Elbaz et al.(2007)]{elb07} Elbaz, D., Daddi, E., Le Borgne, D., et al.\ 2007, \aap, 468, 33 
\bibitem[Font et al.(2008)]{fon08} Font, A.~S., Bower, R.~G., 
McCarthy, I.~G., et al.\ 2008, \mnras, 389, 1619 
\bibitem[G{\'o}mez et al.(2003)]{gom03} G{\'o}mez, P.~L., 
Nichol, R.~C., Miller, C.~J., et al.\ 2003, \apj, 584, 210 
\bibitem[Goto et al.(2003)]{got03} Goto, T., Yamauchi, C., 
Fujita, Y., et al.\ 2003, \mnras, 346, 601 
\bibitem[Hayashi et al.(2010)]{hay10} Hayashi, M., Kodama, 
T., Koyama, Y., et al.\ 2010, \mnras, 402, 1980 
\bibitem[Hopkins(2004)]{hop04} Hopkins, A.~M.\ 2004, \apj, 
615, 209 
\bibitem[Hopkins 
\& Beacom(2006)]{hop06} Hopkins, A.~M., \& Beacom, J.~F.\ 2006, \apj, 651, 142 
\bibitem[Ideue et al.(2009)]{ide09} Ideue, Y., Nagao, T., 
Taniguchi, Y., et al.\ 2009, \apj, 700, 971 
\bibitem[Ideue et al.(2012)]{ide12} Ideue, Y., Taniguchi, Y., 
Nagao, T., et al.\ 2012, \apj, 747, 42 
\bibitem[Ilbert et al.(2009)]{ilb09} Ilbert, O., Capak, P., 
Salvato, M., et al.\ 2009, \apj, 690, 1236 
\bibitem[Ilbert et al.(2010)]{ilb10} Ilbert, O., Salvato, M., 
Le Floc'h, E., et al.\ 2010, \apj, 709, 644 
\bibitem[Kajisawa et al.(2010)]{kaj10} Kajisawa, M., 
Ichikawa, T., Yamada, T., et al.\ 2010, \apj, 723, 129 
\bibitem[Kauffmann et al.(2004)]{kau04} Kauffmann, G., White, 
S.~D.~M., Heckman, T.~M., et al.\ 2004, \mnras, 353, 713 
\bibitem[Koyama et al.(2010)]{koy10} Koyama, Y., Kodama, T., 
Shimasaku, K., et al.\ 2010, \mnras, 403, 1611 
\bibitem[Lacy et al.(2004)]{lac04} Lacy, M., 
Storrie-Lombardi, L.~J., Sajina, A., et al.\ 2004, \apjs, 154, 166 
\bibitem[Lilly et al.(2009)]{lil09} Lilly, S.~J., Le Brun, 
V., Maier, C., et al.\ 2009, \apjs, 184, 218 
\bibitem[Lilly et al.(2007)]{lil07} Lilly, S.~J., Le 
F{\`e}vre, O., Renzini, A., et al.\ 2007, \apjs, 172, 70
\bibitem[Ly et al.(2007)]{ly07} Ly, C., Malkan, M.~A., 
Kashikawa, N., et al.\ 2007, \apj, 657, 738 
\bibitem[Patel et al.(2009)]{pat09} Patel, S.~G., Holden, 
B.~P., Kelson, D.~D., Illingworth, G.~D., 
\& Franx, M.\ 2009, \apjl, 705, L67 
\bibitem[Patel et al.(2011)]{pat11} Patel, S.~G., Kelson, 
D.~D., Holden, B.~P., Franx, M., \& Illingworth, G.~D.\ 2011, \apj, 735, 53 
\bibitem[Poggianti et al.(2008)]{pog08} Poggianti, B.~M., 
Desai, V., Finn, R., et al.\ 2008, \apj, 684, 888 
\bibitem[Poggianti et al.(2010)]{pog10} Poggianti, B.~M., De 
Lucia, G., Varela, J., et al.\ 2010, \mnras, 405, 995 
\bibitem[Sandage et al.(1979)]{san79} Sandage, A., Tammann, 
G.~A., \& Yahil, A.\ 1979, \apj, 232, 352 
\bibitem[Sanders et al.(2007)]{san07} Sanders, D.~B., 
Salvato, M., Aussel, H., et al.\ 2007, \apjs, 172, 86 
\bibitem[Schechter(1976)]{sch76} Schechter, P.\ 1976, \apj, 
203, 297 
\bibitem[Scoville et al.(2007)]{sco07} Scoville, N., Aussel, 
H., Brusa, M., et al.\ 2007, \apjs, 172, 1 
\bibitem[Shioya et al.(2008)]{shi08} Shioya, Y., Taniguchi, 
Y., Sasaki, S.~S., et al.\ 2008, \apjs, 175, 128 
\bibitem[Shioya et al.(2009)]{shi09} Shioya, Y., Taniguchi, 
Y., Sasaki, S.~S., et al.\ 2009, \apj, 696, 546 
\bibitem[Sobral et al.(2011)]{sob11} Sobral, D., Best, P.~N., 
Smail, I., et al.\ 2011, \mnras, 411, 675 
\bibitem[Sobral et al.(2012)]{sob12} Sobral, D., Best, P.~N., 
Matsuda, Y., et al.\ 2012, \mnras, 420, 1926 
\bibitem[Springel et al.(2005)]{spr05} Springel, V., White, 
S.~D.~M., Jenkins, A., et al.\ 2005, \nat, 435, 629 
\bibitem[Stern et al.(2005)]{ste05} Stern, D., Eisenhardt, 
P., Gorjian, V., et al.\ 2005, \apj, 631, 163 
\bibitem[Takahashi et al.(2007)]{tak07} Takahashi, M.~I., 
Shioya, Y., Taniguchi, Y., et al.\ 2007, \apjs, 172, 456
 \bibitem[Tanaka et al.(2004)]{tan04} Tanaka, M., Goto, T., 
Okamura, S., Shimasaku, K., \& Brinkmann, J.\ 2004, \aj, 128, 2677 
\bibitem[Taniguchi et al.(2007)]{tan07} Taniguchi, Y., 
Scoville, N., Murayama, T., et al.\ 2007, \apjs, 172, 9 
\bibitem[Tran et al.(2010)]{tra10} Tran, K.-V.~H., Papovich, 
C., Saintonge, A., et al.\ 2010, \apjl, 719, L126 
\bibitem[Westra et al.(2010)]{wes10} Westra, E., Geller, 
M.~J., Kurtz, M.~J., Fabricant, D.~G., 
\& Dell'Antonio, I.\ 2010, \apj, 708, 534 
\end{thebibliography}
\end{document}